\documentclass{article}

\usepackage[english]{babel}
\usepackage[final]{nips_2017}

\usepackage{graphicx}
\usepackage[utf8]{inputenc} 
\usepackage[T1]{fontenc}    
\usepackage{hyperref}       
\usepackage{url}            
\usepackage{booktabs}       
\usepackage{amsfonts}       
\usepackage{nicefrac}       
\usepackage{microtype}      
\usepackage{epstopdf}

\title{Attention based convolutional neural network for predicting RNA-protein binding sites}

\author{
  Xiaoyong Pan \\
  Department of Medical Informatics\\
  Erasmus Medical Center,\\
  Rotterdam, The Netherlands \\
  \texttt{xypan172436@gmail.com} \\
  \And
  Junchi Yan \\
  IBM Research\\
  Shanghai, China \\
  \texttt{yanesta13@163.com} \\
}

\begin{document}

\maketitle

\begin{abstract}
RNA-binding proteins (RBPs)  play crucial roles in many biological processes, e.g. gene regulation. Computational identification of RBP binding sites on RNAs are urgently needed.
In particular, RBPs bind to RNAs by recognizing sequence motifs. Thus, fast locating those motifs on RNA sequences is crucial and time-efficient for
determining whether the RNAs interact with the RBPs or not. In this study, we present an attention based convolutional neural network, iDeepA, to predict RNA-protein binding sites
from raw RNA sequences. We first encode RNA sequences into one-hot encoding. Next, we design a deep learning model  with a convolutional neural network (CNN)
and an attention mechanism, which automatically search for important positions, e.g. binding motifs, to learn discriminant high-level features for predicting RBP binding sites.
We evaluate iDeepA on publicly gold-standard RBP binding sites derived from CLIP-seq data. The results demonstrate iDeepA achieves comparable performance
with other state-of-the-art methods.
\end{abstract}

\section{Introduction}

RNA-binding proteins (RBPs) take over about 10\% of the eukaryotic proteome and are closely associated with many biological processes [1]. How to identify whether a RNA binds
to a RBP is important for further analyzing the RNAs’ functions. Many experimental technologies have been developed. such as CLIP-seq.
However, they are still time-consuming and high-cost. Thus, computational identification of RBP binding sites are urgently needed.
To this end, many machine learning based methods have been proposed. For example, GraphProt encodes RNA sequences and structures in a graph,
which is further fed into support vector machine to classify bound sites from unbound sites [2]. iONMF integrates multiple sources of data
to predict RBP binding sites using Orthogonal matrix factorization [3].

Recently, deep learning have been successfully developed to predict RNA binding sites. For example, deepnet-rbp applies deep belief network to integrate k-mer
frequency features of sequences
and structures to model RBP targets [4]. DeepBind [5] applies a convolutional neural network (CNN) [6] to identify RBP binding sequence specificity.
iDeep uses multimodal deep learning to integrate different sources of data to infer RBP binding sites and sequence motifs [7].
iDeepS infers sequence and structure motifs simultaneously using a convolutional neural network and long short temporal network [8].
The core of all the above methods is CNN, which demonstrates high accuracy for identifying RBP binding sites.

It is commonly assumed that a RNA sequence that can be bound by a RBP, which contains at least one binding subsequence (motif) of this RBP.
Therefore, it is fairly intuitive to consider putting more attention on this motif subsequence along the RNA sequence. To better model this characteristics of RBP binding sites,
attention mechanism is introduced [9]. Attention mechanism allows deep learning models to focus selectively on only the important features. Deep models augmented with attention mechanisms
have obtained great success on machine translation [9, 10], and computational biology [11].

In this study, we propose an attention-based convolutional neural network model, iDeepA, to predict RBP binding sites from RNA sequences alone. iDeepA combines learned features
from CNNs and two levels of attentions to locate important subsequences.

\section{Method and Materials}
\label{gen_inst}

\subsection{Dataset}
We download RBP binding sites dataset derived from CLIP-seq from GraphProt (\url{http://www.bioinf.uni-freiburg.de/Software/GraphProt}) [2].
It contains 24 experiments of 21 RBPs. For each RBP, it has thousands of bound RNA subsequences with variable length, and almost the same number of negative sequences are selected
with no evidence showing they are bound to this RBP.

\subsection{iDeepA}
\begin{figure}[h]
  \centering
  \includegraphics[width=0.8\linewidth]{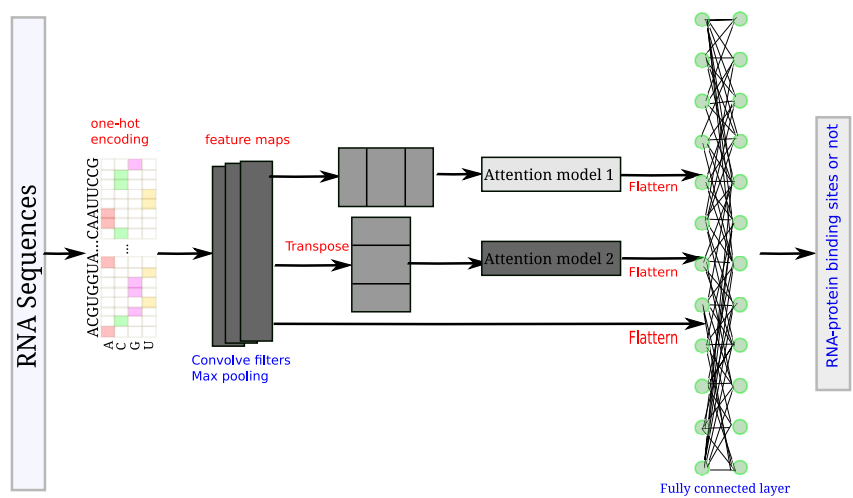}
  \caption{The flowchart of iDeepA. iDeepA first encodes the sequence into one-hot matrix, which is fed into a CNN to output feature maps. Next, we input the last hidden states of the CNN
  to an attention model, and its transposition into another attention model. In the end, the outputs from two attention models and the CNN are combined into two fully connected layers to
  predict RBP binding sites.}
   \label{fig1}
\end{figure}

In this study, we present a CNN based method with attention mechanism to classify RBP bound sites from unbound sites (Figure~\ref{fig1}).
We first encode RNA sequences into one-hot encoding showing the presence of nucleotide A,C,G,U. Then the one-hot encode matrix is fed into a CNN,
which involves convolution, activation, and max-pool operations. The CNN layer preserves the spatial information and output feature maps for subsequent processing.
Inspired by [9, 10], we introduce attention mechanism to further attend differentially to related motifs and locate important positions for predicting RBP binding sites.
We extract three levels of abstract features: 1) The output feature maps from the CNN. 2) The outputs from attention model 1 for sequence dimension, whose input is one copy of
the two-dimensional hidden states from the CNN. 3) The outputs from another attention model 2 for feature map dimension, whose input is transposition of hidden states from the CNN.
For both attention models, we use the same structure with a feedforward neural network as decoder to generate a representation vector.
The output O from an attention model are:
\begin{equation}
 O =  \sum_{t=1}^{T}{h_{t}*\alpha_{t}}
\end{equation}
where $h_{t}$ is hidden state from the CNN and $\alpha_{t}$ is the softmax weight of each hidden state $h_{t}$:
\begin{equation}
\alpha_{t} =  \frac{exp(e_{t})}{\sum_{i=1}^{T}{e_{i}}}
\end{equation}
where $e_{t}$ is generated from the hidden state $h_{t}$ by a feedforward neural network.

By augmenting with the attention mechanism, it learns a soft transformation between the input and output sequences.
Finally, the outputs from CNN layer and two attention models are connected to two fully connected layers. The last layer is the sigmoid layer used to classify the RBP bound sites
from unbound sites. We optimize a categorical entropy loss function using RMSProp [12] with number of epochs 30. iDeepA is implemented using Keras 1.1.2 library \url{https://github.com/fchollet/keras}.

\subsection{Baseline methods}
We compare iDeepA with other state-of-the-art methods, GraphProt, deepnet-rbp, Deepbind and MILCNN. A negative sequence has no any binding site, while a positive sequence
contains at least one binding sites of this RBP. It is intuitive to consider each sequence
as a bag, whose any subsequence is an instance. Inspired by the characteristics, MILCNN first breaks each RNA sequence into multiple overlapping fixed-length subsequence,
each subsequence is an instance and each sequence is a bag of instances. Next, MILCNN trains a CNN under the multiple instance learning framework. 
Multiple instance learning has been used for predicting protein-DNA interactions [14].

\section{Results}
\begin{figure}[h]
  \centering
  \includegraphics[width=0.8\linewidth]{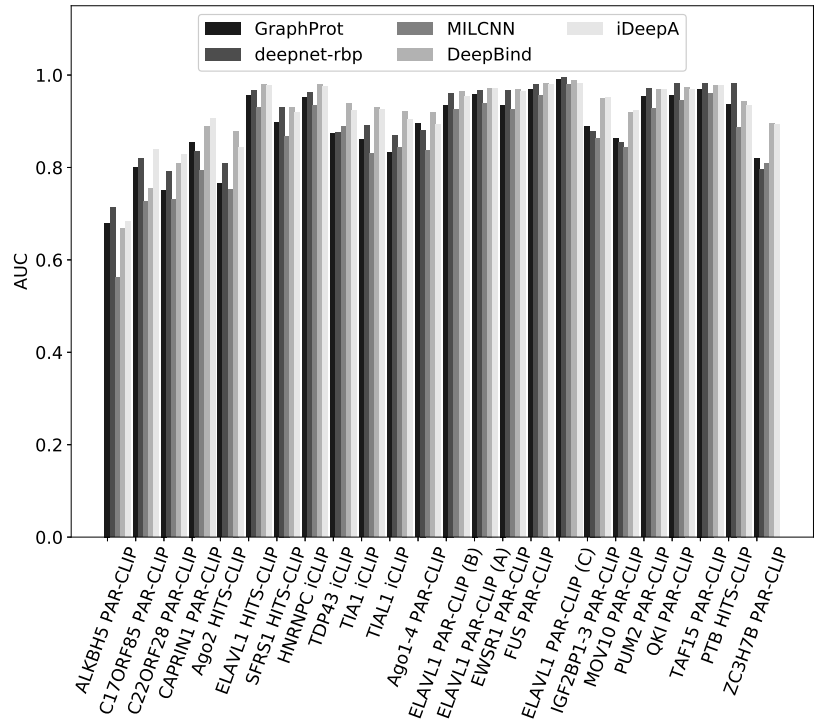}
  \caption{The AUCs of different methods for predicting RBP binding sites. The AUCs of GraphProt and deepnet-rbp are taken from original papers, other three methods are ran on
  the same training and testing set with similar CNN network.}
   \label{fig2}
\end{figure}

GraphProt, deepnet-rbp, MILCNN, DeepBind and iDeepA achieve the average AUC 0.887, 0.902, 0.861, 0.921 and 0.921 across 24 experiments (Figure~\ref{fig2}), respectively.
iDeepA and DeepBind yield similar average AUC, which is higher than other three methods. In addition, iDeepA improves some RBPs with small training set
on that DeepBind does not achieve high AUC.
For example, iDeepA obtains an AUC of 0.839 for C17ORF85 with only 4000 training samples, which is an increase by 11\% compared to an AUC 0.755 of DeepBind. The results indicates
introducing attention mechanism can enhance the learning ability on small dataset than DeepBind and it is fast to focus on important subsequences. 
However, introducing attention mechanism does not improve the performance on those RBPs with large number of training samples, it is possible because 
feeding more samples into model training can make the model to converge to the same optimum model. 
In addition, MILCNN yields lower performance than other methods,
it maybe because that training RNA sequences are themselves subsequence anchored at the peak center derived from CLIP-seq, breaking them into subsequence
may also break the binding sites.

\label{headings}

\section{Conclusion}
In this study, we present an attention-based CNN method to predict RBP binding sites. Our method iDeepA yields comparable performance with other state-of-the-art methods.
However, we still do not further investigate whether the attention can be used to identify interpretable motifs.
In future work, we expect to obtain more interpretablitity of iDeepA and comprehensively evaluate iDeepA on larger dataset with more RBPs.
\section*{References}

\small
[1] Ray,D., Kazan,H., \textit{et~al}. (2013) A compendium of RNA-binding motifs for decoding gene regulation. {\it Nature.} {\bf 499}, 172-7. doi: 10.1038/nature12311.

[2] Maticzka,D., Lange,S.J.,Costa,F., Backofen,R. (2014) GraphProt: modeling binding preferences of RNA-binding proteins. {\it Genome Biol.} {\bf 15,} R17.  doi: 10.1186/gb-2014-15-1-r17.

[3] Stražar,M., Žitnik,M., Zupan,B., Ule,J., Curk,T. (2016) Orthogonal matrix factorization enables integrative analysis of multiple RNA binding proteins. {\it Bioinformatics.} {\bf 32}, 1527-35. doi: 10.1093/bioinformatics/btw003.

[4] Zhang,S., Zhou,J., Hu,H., Gong,H., Chen,L., Cheng,C., Zeng,J. (2015) A deep learning framework for modeling structural features of RNA-binding protein targets. {\it Nucleic Acids Res.} {\bf 44}, e32. doi: 10.1093/nar/gkv1025

[5] Alipanahi, B., et al. (2015) Predicting the sequence specificities of DNA- and RNA-binding proteins by deep learning, {\it Nature biotechnology}, {\bf 33}, 831-838.

[6] LeCun,Y., Léon,B., Yoshua,B.\ \&Patrick,H. (1998) Gradient-based learning applied to document recognition. {\it Proceedings of the IEEE. 1998.} {\bf 86}, 2278-2324.

[7] Pan,X.\ \& Shen,H.B. (2016) RNA-protein binding motifs mining with a new hybrid deep learning based cross-domain knowledge integration approach. {\it BMC Bioinformtics.} {\bf 18,} 136.

[8] Pan,X., Rijnbeek,P. , Yan,J. \ \&  Shen,H.B. (2017) Prediction of RNA-protein sequence and structure binding preferences using deep convolutional and recurrent neural networks. biorxiv 146175.

[9] Bahdanau, D., Cho, K. \ \& Bengio, Y. (2014) Neural machine translation by jointly learning to align and translate, arXiv preprint arXiv:1409.0473.

[10] Vaswani,A. \textit{et~al}. (2017) Attention Is All You Need. arXiv:1706.03762.

[11]  Wang,D. \textit{et~al}. (2017) MusiteDeep: a Deep-learning Framework for General and Kinase-specific Phosphorylation Site Prediction. {\it Bioinformatics.} btx496, https://doi.org/10.1093/bioinformatics/btx496

[12] Tieleman,T.\ \& Hinton,G.E. (2012) Lecture 6.5 - rmsprop: Divide the gradient by a run-ning average of its recent magnitude. {\it COURSERA: Neural Networks for Machine Learning.} {\bf 4,} 2.

[13] Pan,S.J.\ \&Yang,Q. (2010) A Survey on Transfer Learning, IEEE Transactions on Knowledge and Data Engineering (IEEE TKDE), 22(10):1345-1359.

[14] Gao,Z. \ \& Ruan,J. (2017) Computational modeling of in vivo and in vitro protein-DNA interactions by multiple instance learning. {\it Bioinformatics.} {\bf 33}:2097-2105.

\end{document}